\documentclass[12pt]{article}
\usepackage[authoryear]{natbib}
\usepackage{amssymb,amsthm,epsfig}
\usepackage[centertags]{amsmath}
\usepackage{bm}
\usepackage{lineno}

\newtheorem{teorema}{Theorem}

\newtheorem{proposition}{Proposition}

\renewcommand{\baselinestretch}{1.3}

\begin{document}
\vskip 0.3 cm 
\begin{center}
{\bf\large Vector Autoregressive Models With Measurement Errors\\
for Testing Ganger Causality} 
\end{center}\vskip 0.3 cm

\centerline{\bf Alexandre G. Patriota$^*$, \ Jo\~ao R. Sato$^{**}$
\ and  \ Betsab\'e G. Blas Achic$^*$}

 \vskip 0.3 cm
\centerline{$^*$Departamento de Estat\'{\i}stica, Universidade de
S\~{a}o Paulo - SP - Brasil}

\centerline{Caixa Postal 66281 - CEP 05314-970, S\~{a}o Paulo - SP
- Brasil}

\centerline{$^{**}$ Institute of Radiology - Hospital das
Cl\'inicas, S\~{a}o Paulo - Brasil}

\centerline{CEP 05403-001, S\~{a}o Paulo - SP - Brasil}
\begin{abstract}
This paper develops a method for estimating parameters of a vector autoregression (VAR) observed 
in white noise. The estimation method assumes the noise variance matrix is known and does not require
any iterative process. This study provides consistent estimators and shows the asymptotic distribution of the 
parameters required for conducting tests of Granger causality. Methods in the 
existing statistical literature cannot be used for testing Granger causality, since under the null hypothesis
the model becomes unidentifiable. Measurement error effects on parameter estimates were evaluated by using 
computational simulations.
The results show that the proposed approach produces empirical false positive rates
 close to the adopted nominal level (even for small samples) and
has a good performance around the null hypothesis.
 The applicability and usefulness of the proposed approach are illustrated
using a functional magnetic resonance imaging dataset.\\

\end{abstract}

\section{\protect \centering INTRODUCTION}
Multivariate time series modeling is an important component
for the quantitative assessment of relationships between variables in
many applied areas. This issue is essential in financial applications, 
for example, enabling optimal portfolio
allocation, setting trading strategies over sectors of the
market, or exchanging rates \citep{sims,ni}. In addition, the vector autoregressive
model (VAR) is widely used in many fields such as economics \citep{granger69}, geophysics \citep{liu05},
bioinformatics \citep{fujita07a} and neuroscience \citep{goebel03}.

The main reasons for the attractiveness of the VAR model in applied areas
are its simplicity and relation with the concept of Granger causality \citep{granger69}. 
Granger causality has become a prominent concept in connectivity
networks modeling, because it provides inferences about the direction of
information flow between different time series. Several studies in biological
systems emphasize the importance of identification and
description of gene regulator networks \citep{gottesman94,katoh07},
mainly in the study of tumors or structural diseases.
\cite{mukhopadhyay07,fujita07a,fujita07b} introduced the utilization
of VAR-based models to study these issues by applying these models to
gene expression datasets. In Neuroscience, the \emph{functional integration} theories highlight
that brain functions heavily depend on neural connectivity networks \citep{cohen01}.
Several neuroimaging studies \citep{goebel03,sato06,abler06}
suggested that VAR models and Granger causality are suitable to identify
the information flow between neural structures. Nevertheless, it is well
known that most biological measurements are subject to error, since the
precision of acquisition equipments is never absolute. Actually, this
limitation is present in most studies involving experimental data, such as
chemistry, physics, biometrics, etc.

Although technically incorrect, the most common procedure is
simply to ignore the measurement errors, i.e.: assume that the variables of interest
are the observed variables. It is important to highlight that this assumption
has serious implications. The utilization of conventional VAR model in this
case would not identify correctly the relationships
between the variables of interest (latent variables). It happens because the model
white noise will not be independent which leads to misestimations of the model
parameters. The usual assumption is acceptable when the errors are
negligible. However, it is known that due to acquisition processes limitations,
the measurement errors in biology (e.g.: gene expressions or brain signals) are
not negligible. In these cases, the utilization of conventional VAR models may
result in biased parameter estimation and as a consequence, unreliable Granger
causality detection.

In the following, we define the usual VAR model 
\citep[for a more detailed description, see for instance,][]{lutkepohl}. Let  $\bm{z}_t =(z_{1t},\ldots, z_{pt})^{\top}$
denotes a $(p\times 1)$ vector of time series variables. The usual
 VAR(r) model has the form
\begin{equation}\label{var}
\bm{z}_t=\bm{a}+\bm{B}_1\bm{z}_{t-1}+\ldots +\bm{B}_r \bm{z}_{t-r}
+ \bm{q}_t,\,\,\,t=1,\cdots,n
\end{equation}
where $n$ is the sample size, $\bm{B}_j$ for $j=1,\ldots,r$ are $(p\times p)$ coefficient
matrices and $\bm{q}_t$ is a $(p\times 1)$ unobservable zero mean
white noise vector process with covariance matrix $\bm{\Sigma}$.
 For convenience, we consider that $ \bm{z}_l=\bm{0}$ for all $l \leq 0$.
We are assuming throughout this paper that model (\ref{var})
 satisfies the stability condition defined in \cite{lutkepohl} on page 12.
 Therefore, under stationarity conditions, the mean and the autocovariance
function are given, respectively, by
\[
E(\bm{z}_t) = \bm{\mu_z} = \bigg(\bm{I}_p - \sum_{j=1}^r
\bm{B}_j\bigg)^{-1}\bm{a},\]
\[\bm{\gamma}(h) = E[(\bm{z}_t -\bm{\mu_z})(\bm{z}_{t-h} -\bm{\mu_z})^{\top}]=\sum_{j=1}^r \bm{B}_j \bm{\gamma}(h-j), \quad \mbox{for} \ \  h = 1,2,3, \ldots
\] and
\[
\bm{\gamma}(0) = \sum_{j=1}^r \bm{B}_j \bm{\gamma}(-j) + \bm{\Sigma}
\]
 where $\bm{I}_p$ denotes the $p\times p$ identity matrix and $\bm{\gamma}(-j)=\bm{\gamma}(j)^{\top}$.

Model (\ref{var}) can be written in short as
\begin{equation}\label{varCOL}
\bm{z}_t=\bm{a}+\bm{B}\bm{z^*}_{t-1} + \bm{q}_t,\,\,\,t=1,\cdots,n
\end{equation}
 where $\bm{B} = (\bm{B}_1 \ \bm{B}_2 \ \ldots \ \bm{B}_r)$ is a $p\times pr$ matrix
 and $\bm{z^*}_{t-1} = (\bm{z}_{t-1}^{\top}, \bm{z}_{t-2}^{\top},
\ldots, \bm{z}_{t-r}^{\top})^{\top}$.

Therefore, if the white noise has normal distribution, the conditional Maximum
Likelihood (ML) estimators of ${\bm{a}}$, ${\bm{B}}$ and  ${\bm{\Sigma}}$ are
 equal to the ordinary least squares estimators. They are given, respectively, by
\begin{equation}\label{Vest}
\widehat{\bm{a}}_{_{ML}} = \bar{\bm{z}}_{t} -
\widehat{\bm{B}}_{_{ML}}\bar{\bm{z^*}}_{t-1}, \quad \widehat{\bm{B}}_{_{ML}}
= (\bm{S}_{\bm{z^*}_{t-1}}^{-1}
\bm{S}_{\bm{z^*}_{t-1}\bm{z}_t})^{\top}\quad \mbox{and} \quad
\widehat{\bm{\Sigma}}_{_{ML}} = n^{-1}\sum_{i=1}^n
\widehat{\bm{q}}_i\widehat{\bm{q}}_i^{\top}
\end{equation} where $\bar{\bm{z^*}}_{t-1} = n^{-1}\sum_{i=1}^n \bm{z^*}_{i-1}$, $\bar{\bm{z}}_t = n^{-1}\sum_{i=1}^n \bm{z}_i$,
 $\widehat{\bm{q}}_i = \bm{z}_i - \widehat{\bm{a}}_{_{ML}} - \widehat{\bm{B}}_{_{ML}}\bm{z^*}_{i-1}$,
$\bm{S}_{\bm{z^*}_{t-1}}=n^{-1}\sum_{i=1}^n(\bm{z^*}_{i-1}-\bar{\bm{z^*}}_{t-1}){\bm{z^*}_{i-1}}^{\top}$
and $\bm{S}_{\bm{z^*}_{t-1}\bm{z}_t}
=n^{-1}\sum_{i=1}^n(\bm{z^*}_{i-1}-\bar{\bm{z^*}}_{t-1}){\bm{z}_{i}}^{\top}$.

The consistence of those conditional ML estimators is assured under the stationary
conditions \citep[see][for further details]{lutkepohl}. The consistence is shown
using the fact that

\[\bar{\bm{z}}_{t} \stackrel{\mathcal{P}}{\longrightarrow}
\bm{\mu_z}, \quad \bar{\bm{z^*}}_{t-1}
\stackrel{\mathcal{P}}{\longrightarrow} \bm{\mu_{z^*}} =
\bm{1}_r\otimes \bm{\mu_z},\quad
 \bm{S}_{\bm{z^*}_{t-1}}\stackrel{\mathcal{P}}{\longrightarrow} \bm{\Gamma}_r(0)\quad \mbox{and} \quad
\bm{S}_{\bm{z^*}_{t-1}\bm{z}_t}\stackrel{\mathcal{P}}{\longrightarrow}
\bm{\Gamma}_r(0)\bm{B}^{\top}\] where
``$\stackrel{\mathcal{P}}{\longrightarrow}$" denotes convergence in
probability when the sample size increases, $\otimes$  denotes the Kronecker product,
$\bm{1}_r$ is a $r-$dimensional column vector of ones, and  the
covariance function of $\bm{z^*}_{t-1}$ is given by
\begin{eqnarray*}
\bm{\Gamma}_r(h) &=& E[(\bm{z^*}_{t-1} - \bm{\mu_{z^*}}
)(\bm{z^*}_{t-h-1} - \bm{\mu_{z^*}} )^{\top}]\\
&=& \left[
\begin{array}{cccc}
\bm{\gamma}(h)     & \bm{\gamma}(h+1)   & \ldots & \bm{\gamma}(h+r-1)\\
\bm{\gamma}(h-1)   & \bm{\gamma}(h)     & \ldots & \bm{\gamma}(h+r-2)\\
\vdots             & \vdots             & \ddots & \vdots \\
\bm{\gamma}(h-r+1) & \bm{\gamma}(h-r+2) & \ldots & \bm{\gamma}(h)
\end{array}\right].\end{eqnarray*}

As described previously, VAR modeling is commonly applied for detecting Granger
causality relationships. The basic idea of Granger causality is the evaluation of
temporal information founded on the assumption that the cause always precedes its
effect \citep{granger69}. Let $x_t$ and $y_t$ be two time series. From the statistical
perspective, $x_t$ is said to Granger-cause $y_t$ if the prediction error of $y_t$,
conditioning on the past values of both series, is less than considering solely the
past values of $y_t$. In other words, the past values of $x_t$ contains relevant
information to improve the predictions of $y_t$. Note that Granger causality concept
is not equivalent to classical aristotelian causality, since the former is based solely
on prediction errors. However, due to its simplicity, it may be applied to identify
possible effective causalities.

One possible approach of using VAR models for Granger causality detection is by performing
statistical tests on $\bm{B}_j$'s coefficients. Considering $y_t$ equation, if
there is at least one coefficient multiplying the past values of $x_t$ which is not equal
to zero, then $x_t$ is said to Granger-cause $y_t$. Thus, this procedure involves the
estimation of $\bm{{B}}_j$, their respective covariance matrices, and the application
of hypothesis testing.

In general, many physical, biological and chemical variables have the
measurement process subject to random effects and it is very common
analyze them by using models assuming that these measurement errors are
negligible. It may bring up undesirable features as biased estimates as
well as their standard errors  and, as a consequence, dangerously false
confidence intervals and hypotheses testing will often be obtained using
such approach. Thus, it is necessary to consider the measurement error on the
modeling of these type of time series.

In this paper, we study a VAR model with main concern on including measurement 
errors. Let $\bm{z}_t$ be the true variable that is not directly  observed, 
instead a substitute variable $\bm{Z}_t$ is observed which has an additive structure given by

\begin{equation}\label{erro}
\bm{Z}_t= \bm{z}_t+\bm{e}_{t},\,\,\,t=1,\cdots,n
\end{equation}
 where
$\bm{Z}_t=(Z_{1t},Z_{2t},\cdots,Z_{pt})^{\top}$ is the observed
vector and $\bm{e}_{t}=(e_{1t},e_{2t},\cdots,e_{pt})^{\top}$ is
the measurement error vector with mean zero and variance-covariance matrix
$\bm{\Sigma}_e$. In most cases, if the usual conditional
ML estimator is adopted for the observations subject to errors, i.e.,
replacing $\bm{z}_t$ with $\bm{Z}_t$ in the equation (\ref{var}), the
estimator of $\bm{B}$ will be biased (as can be seen in (\ref{olsp})). 
Therefore, in order to overcome this limitation
the measurement errors should be included in the estimation procedure.
 Nevertheless, the model (\ref{var}) with the equation (\ref{erro}) is
not identifiable, since the covariance matrices of $\bm{q}_t$ and $\bm{e}_t$
are confounded when $\bm{B} = \bm{0}$. It is easy to see that in the univariate AR(1),
note that when $r=p=1$ and $b=0$ we have: $Z_i = a + q_i + e_i$ with $\mbox{E}(Z_i) = a$, 
$\gamma(0) = \sigma^2 + \sigma_e^2$ and $\gamma(h) = 0$ for all $h\neq 0$. It is impossible
to estimate $\sigma^2$ and $\sigma^2_e$ separately by observing only $Z_1, \ldots, Z_n$.
 This problem can be avoided by using previous knowledge about
the variance of $\bm{e}_t$.

This paper is organized as follows. Section 2 proposes consistent
estimators for the VAR model with measurement errors and also presents the
asymptotic distribution of the estimator of the elements of $\bm{B}$.
In Section 3, simulation studies are undertaken to investigate some aspects
of the proposed estimators (rejection rates for a test of hypothesis, biases and mean
square errors) also it is verified the impact by erroneously considering
the usual model. We applied the models in a functional magnetic resonance
imaging dataset in Section 4 and we
finish the paper with conclusions and remarks in Section 5.

\section{\protect \centering VAR WITH MEASUREMENT ERRORS}
In the presence of measurement errors, the conventional
ML estimation of VAR models produces biased estimators and they can lead to
wrong statistical inference  \citep[see][in which it is found a discussion
over errors-in-variables in regression models]{fuller}. There are some studies about measurement errors
in times series \citep[e.g.,][]{geweke,aigner}. Those studies use Kalman filtering 
methodology and an Expectation and Maximization algorithm
that requires intensive iterative procedures. \cite{maravall1}
have provided a careful expose of the identifiability of some time
series models with errors in variables. \cite{beck} describes approaches based on
state space modeling and Kalman filtering and demonstrates the
usefulness of these tools in dynamic models. \cite{kellstedt} show the efficiency
gains adopting an errors-in-variables model, and the precision of
Kalman filter estimates in the face of autocorrelation. These
measurement techniques have been applied to a variety of
substantive problems, including dynamic representation, social
problems (such as racial inequality), monetary policy and public
entrepreneurship ({\bf Citar}).

These state space models can be attractive alternatives to conventional VAR modeling.
However, in practice, the implementation of the estimators are not described in
analytical form, but by interactive algorithms or numerical optimization solutions.
In addition, the derivation of estimators convergence, standard errors, consistence
and asymptotic distribution may be complex in these cases. 
In \cite{Shumway}, the section on state space methods shows an alternative procedure 
for how to estimate $\bm{B}$, $\bm{\Sigma}$ and $\bm{\Sigma}_e$ under model (\ref{var}) 
with the error equations (\ref{erro}), using the EM algorithm. \cite{hannan2003} proposed 
another iterative procedure to estimate these parameters. Nevertheless, as the main goal
of this paper is to test Granger causality, these approaches can not be used, since the model
becomes unidentifiable under the hypothesis $\bm{B}=\bm{0}$.

In this study, we provide simple and closed forms for the estimators when $\bm{\Sigma}_e$ is known, 
which allows the direct derivation of their respective asymptotic properties. Since the main concern
of several practical applications is Granger causality testing, this information is
essential to data analysis. In this section, the main concern is the parameter estimation and its asymptotic properties. 
Theorem \ref{EstHomo} states consistent estimators
for the model parameters and Theorem \ref{AsympEstHomo}
establishes the asymptotic distribution for the estimator
 of $\mbox{vec}(\bm{B}^{\top})$ given in Theorem \ref{EstHomo}, where
$\mbox{vec}(\bm{C})$ is an operator that heaps the columns of the matrix $\bm{C}$.

The methodology presented in this section is based on correcting the asymptotic bias
of conventional ML estimator caused by the measurement error effect. The outcome is a consistent
estimator with good asymptotic properties such as normality. The estimators and the
asymptotic covariance matrix for the proposed estimator of $\mbox{vec}(\bm{B}^{\top})$
 are computed easily and no iterative procedure is required. We must remark that those
estimators are not the conditional ML estimators nor the ML estimators taking into account
the measurement errors which are very complicated to reach by maximizing the likelihood,
 even under normality of the errors.

\begin{teorema}\label{EstHomo}
If $\bm{e}_t\sim \mathcal{N}(\bm{0}, \bm{\Sigma}_{e})$ with $\bm{\Sigma}_{e}$ known.
Then, the parameters of the model (\ref{var}) under measurement errors as in (\ref{erro}) have
consistent estimators given by

\begin{equation}\label{Beta}
\widehat{\bm{a}}= \bar{\bm{Z}}_t-\widehat{\bm{B}}\bar{\bm{Z^*}}_{t-1}, \quad
\widehat{\bm{B}}=\bigg[(\bm{S}_{\bm{Z^*}_{t-1}}- \bm{I}_r\otimes\bm{\Sigma}_{e})^{-1}\bm{S}_{\bm{Z^*}_{t-1}\bm{Z}_t}\bigg]^{\top}
\end{equation} and

\[
\widehat{\bm{\Sigma}} = n^{-1} \sum_{i=1}^{n} (\bm{Z}_i - \widehat{\bm{a}} - \widehat{\bm{B}}\bm{Z^*}_{i-1})(\bm{Z}_i- \widehat{\bm{a}} - \widehat{\bm{B}}\bm{Z^*}_{i-1})^{\top} -\bm{\Sigma}_e - \widehat{\bm{B}}(\bm{I}_r\otimes\bm{\Sigma}_e)\widehat{\bm{B}}^{\top}
\] where $\bar{\bm{Z^*}}_{t-1} = n^{-1}\sum_i \bm{Z^*}_{i-1}$, $\bar{\bm{Z}}_t = n^{-1}\sum_i \bm{Z}_i$,
$\bm{S}_{\bm{Z^*}_{t-1}}=n^{-1}\sum_{i}(\bm{Z^*}_{i-1}-\bar{\bm{Z^*}}_{t-1}){\bm{Z^*}_{i-1}}^{\top}$ and
$\bm{S}_{\bm{Z^*}_{t-1}\bm{Z}_t} =n^{-1}\sum_{i}(\bm{Z^*}_{i-1}-\bar{\bm{Z^*}}_{t-1}){\bm{Z}_{i}}^{\top}$.
\end{teorema}

The proof of Theorem \ref{EstHomo} can be found in Appendix
\ref{A}. Notice that, if $\bm{\Sigma}_e = \bm{0}_{p\times p}$, that is,
when there is no measurement error, then
the estimators of Theorem \ref{EstHomo} become the conditional ML
estimators presented in (\ref{Vest}). Also, it can be seen that
the conditional ML estimator of $\bm{B}$ from the model (\ref{var}),
without considering the errors (\ref{erro}), is given
by
\[\widehat{\bm{B}}_{ML}=\bigg[\bm{S}_{\bm{Z^*}_{t-1}}^{-1}\bm{S}_{\bm{Z^*}_{t-1}\bm{Z}_t}\bigg]^{\top},
\] which is not consistent, since

\begin{equation}\label{olsp}\widehat{\bm{B}}_{ML}\stackrel{\mathcal{P}}{\longrightarrow}\bm{B}
[\bm{I}_{pr}+(\bm{I}_r\otimes\bm{\Sigma}_e)\bm{\Gamma}_r(0)^{-1}]^{-1}.
\end{equation}

The main steps to demonstrate (\ref{olsp}) is given in Appendix \ref{A}, in which is
sufficient to compute the limit of $\bm{S}_{\bm{Z^*}_{t-1}}$ and
$\bm{S}_{\bm{Z^*}_{t-1}\bm{Z}_t}$. The quantity  $\bm{S}_{\bm{Z^*}_{t-1}}$ has
two sources of variations, one that refers to the unobservable variable $\bm{z^*}_{t-1}$
and another one that refers to the measurement error. If the measurement error
is huge and the sample size is not large enough, the quantity
$(\bm{S}_{\bm{Z^*}_{t-1}}- \bm{I}_r\otimes\bm{\Sigma}_{e})$
may not be positive definite and the estimator $\widehat{\bm{B}}$, presented
in (\ref{Beta}), will be inadmissible. If the quantity
$(\bm{S}_{\bm{Z^*}_{t-1}}- \bm{I}_r\otimes\bm{\Sigma}_{e})$
 has at least one eigen value close to zero the estimator $\widehat{\bm{B}}$,
presented in \ref{Beta}, will be unstable (because the computation of a matrix inverse
requires all eigen values to be different from zero). If the matrix $\bm{\Sigma}_e$
is well specified, one way to avoid such inadmissibility and instability is
increasing the sample size.

In many practical applications, there is some interest on testing some
elements of the matrix $\bm{B}$ (e.g., the so called Granger causality
test). However, the exact distribution of $\mbox{vec}(\widehat{\bm{B}}^{\top})$
is hard to compute. Thus, one can use its asymptotic distribution to build
confidence regions and hypothesis testing as an approximation when the sample
size is finite. The Theorem below gives us the asymptotic distribution of
$\mbox{vec}(\widehat{\bm{B}}^{\top})$.

\begin{teorema}\label{AsympEstHomo}
If $\bm{e}_{t} \sim \mathcal{N}(\bm{0}, \bm{\Sigma}_e)$ with
$\bm{\Sigma}_{e}$ known and $\mbox{E}(q_{ij_1}q_{ij_2}q_{ij_3}q_{ij_4}) < \infty$
for all $j_1, j_2, j_3, j_4 \in \{1,\ldots, p\}$, where $q_{ij}$ is the $j^{th}$ element of
$\bm{q}_i$. Then, the asymptotic distribution of
$\mbox{vec}(\widehat{\bm{B}}^{\top})$ obtained in Theorem \ref{EstHomo}
 is given by

 \begin{equation}\label{teoH}\sqrt{n}(\mbox{vec}(\widehat{\bm{B}}^{\top})-\mbox{vec}(\bm{B}^{\top}))\stackrel{D}{\longrightarrow}
 \mathcal{N}(\bm{0},\bm{\Phi}),
\end{equation}
where the $p^2r\times p^2r$ matrix $\bm{\Phi}$ is given by

\begin{equation*}\bm{\Phi}=\bm{\Sigma}_{\vartheta}\otimes\bm{\Gamma}_r(0)^{-1}+(\bm{I}_p\otimes\bm{\Gamma}_r(0)^{-1})\bm{A}_r(\bm{I}_p\otimes\bm{\Gamma}_r(0)^{-1})
 \end{equation*}
where

\begin{eqnarray*}
\bm{A}_r& =& \bm{\Sigma}_{\vartheta} \otimes (\bm{I}_r \otimes \bm{\Sigma}_e) + \bm{B}^{\top} \otimes  [\bm{\Sigma}_e\bm{B} (\bm{I}_r \otimes \bm{\Sigma}_e)] +\\
&&  - \sum_{h=1}^{r}\bigg\{( \bm{B}_{h} \bm{\Sigma}_{e})\otimes\bm{\Gamma}_r(h) + (\bm{\Sigma}_{e}\bm{B}_{h}^{\top})\otimes\bm{\Gamma}_r(-h)\bigg\} +\\
&& +  \sum_{h=1-r}^{r-1} [\bm{B}(\bm{J}_{-h} \otimes \bm{\Sigma}_e)\bm{B}^{\top}]\otimes \bm{\Gamma}_r(h).
\end{eqnarray*}
and $\bm{\Sigma}_{\vartheta}=\bm{\Sigma}+\bm{\Sigma}_{e}+\bm{B}(\bm{I}_r\otimes\bm{\Sigma}_{e})\bm{B}^{\top}$,
where $\bm{J}_{l}$ is a ($r \times r$) matrix of zeros with one's in the $|l|^{th}$ diagonal above (below)
the main diagonal if $l>0$ ($l<0$) and $\bm{J}_0$ is a ($r \times r$) matrix of zeros.

\end{teorema}

The proof of Theorem \ref{AsympEstHomo} can be seen in Appendix \ref{B}.
For all $r$ and $\bm{\Sigma}_{e}=0$ we have $\bm{\Phi}=\bm{\Sigma}\otimes
\bm{\Gamma}_r(0)^{-1}$, as given in \cite{lutkepohl}. The normal distribution
assumption for the measurement error is
required to compute the expectation of polynomial functions (until
forth degrees) of the elements of $\bm{e}_t$. Notice that, if $r=1$
we have the $\mbox{VAR}(1)$ model and the asymptotic covariance
simplifies to

\[
\bm{\Phi}=\bm{\Sigma}_{\vartheta}\otimes\bm{\gamma}(0)^{-1}+(\bm{I}_p\otimes \bm{\gamma}(0)^{-1})\bm{A}_1(\bm{I}_p \otimes \bm{\gamma}(0)^{-1})
\] where

\[
\bm{A}_1 = \bm{\Sigma}_{\vartheta}\otimes\bm{\Sigma}_{e}+\bm{B}^{\top}\otimes (\bm{\Sigma}_{e}\bm{B}\bm{\Sigma}_{e})- [(\bm{B} \bm{\Sigma}_{e})\otimes (\bm{\gamma}(0)\bm{B}^{\top})  + (\bm{\Sigma}_{e}\bm{B}^{\top})\otimes(\bm{B}\bm{\gamma}(0))].
\]

The $i^{th}$ element of $\mbox{vec}(\widehat{\bm{B}}^{\top})$,
 is asymptotically normally distributed with standard error
given by the square root of $i^{th}$ diagonal element of
$\bm{\Phi}$. Thus, we can obtain hypotheses tests on the
individual coefficients, or more general form of contrasts

\[H_0: \bm{C}\mbox{vec}(\bm{B}^{\top}) = \bm{d} \qquad \mbox{\emph{Versus}}
\qquad H_1: \bm{C}\mbox{vec}(\bm{B}^{\top}) \neq \bm{d},\] which involves
 coefficients across  different equations of the VAR model.
Thus, Granger causality testing can be carried out by adequately specifying this
contrasts matrix. An illustrative example is the case of series $x_t$ and $y_t$,
in which we are interested in evaluating the Granger causality from $x_t$ to $y_t$
in an $r$-order VAR model. The matrix $\bm{C}$ has $r$ rows, one for each coefficient
related to the past values of $x_t$ in the $y_t$ equation. Considering that each
column of $\bm{C}$ refers to each VAR coefficient, the contrast matrix is specified
by simply setting 1 to the cell at the respective column and row for the $x_t$
coefficients in $y_t$ equation. This may be tested
using the Wald-type statistic conveniently expressed as

\begin{equation}\label{Wald}
n(\bm{C}\mbox{vec}(\widehat{\bm{B}}^{\top})-\bm{d})^{\top}\big[\bm{C}\bm{\Phi}\bm{C}^{\top}\big]^{-1}(\bm{C}\mbox{vec}(\widehat{\bm{B}}^{\top})-\bm{d})
\end{equation}

Under the null hypotheses, (\ref{Wald}) has a $\chi^2(m)$
distribution in the limit, where $m = rank(\bm{C})$ gives the number of linear
restrictions.

The above study can also be developed to the intercept model estimator, it can be
found by applying the delta method \citep{lehcas98} in the asymptotic distribution
of ($\bar{\bm{Z}}_t^{\top}$, $\bar{\bm{Z}}_{t-1}^{\bm{*}\top}$, $\mbox{vec}(\widehat{\bm{B}}^{\top})^{\top}$),
since  $\widehat{\bm{a}} = \bar{\bm{Z}}_t - (\bm{I}\otimes \bar{\bm{Z}}_{t-1}^{\bm{*}\top})\mbox{vec}(\widehat{\bm{B}}^{\top})$. Although, this asymptotic distribution is important to test hypotheses regarding
the  model intercept, it is outside the main scope of this article and does not have any impact
on the Granger causality, for this reason we skip it.

\section{\protect \centering SIMULATION RESULTS}
In this section we conduct some simulation studies in order to
evaluate the adequacy of the asymptotic distribution of
$\mbox{vec}(\widehat{\bm{B}}^{\top})$ for small and moderate
samples sizes. Computations were
performed using the software R (www.r-project.org). 

For each setup of parameters and sample sizes, we considered $15{,}000$ Monte
Carlo samples generated from a VAR(1) model with measurement errors, given by

\begin{eqnarray}\label{m1}
{z_{1,t} \choose z_{2,t}} &=& {a_1\choose a_2} + \left[
\begin{array}{cc}
b_{11} & b_{12}\\
b_{21} & b_{22}
\end{array}
\right]{z_{1,t-1} \choose z_{2,t-1}} + { q_{1t}\choose q_{2t}},\\
{Z_{1,t} \choose Z_{2,t}} &=&{z_{1,t} \choose z_{2,t}} + {
e_{1t}\choose e_{2t}}.
\end{eqnarray}

In all samples, we have considered the following setup of
parameters: $a_1=a_2 = 1$, $b_{11}=b_{22} = 0.5$,

\[
\bm{\Sigma} = \left[
\begin{array}{cc}
10 & 5 \\
5 & 5
\end{array}
\right],
\] where the vector parameters values of ($b_{12},b_{21}$) were
the values of the set $\{(b_{12},b_{21}); b_{12}\in S \mbox{ and }
b_{21} \in S\}$, where $S=\{-0.4, -0.2, 0.0, 0.2, 0.4\}$,  the
variance of the measurement error $\bm{e}_t$ was $\bm{\Sigma}_e =
2 \bm{I}_2$, and the size samples $n=50,100,250,500$.

The rejection rates of the hypothesis $H_0 : b_{12} =b_{21} = 0$
(i.e., $z_{2,t-1}$ does not help to explain $z_{1,t}$ and $z_{1,t-1}$
does not help to explain $z_{2,t}$) are shown in Table \ref{tab:n}, in
which the test sizes are the rejection rates under the null hypothesis
(that appears in bold). The Wald-type
statistics (\ref{Wald}) is used at  5\% nominal level. From this table
we have that, the test sizes from the proposed model
are closer to the nominal level (5\%) as compared to the usual approach
for all sample sizes. Furthermore, when $n$ increases the test sizes
for the usual model also increase and, consequently, they do not converge to
the adopted nominal level. This is a somewhat expected behavior
because the usual approach produces biased estimates and
 standard errors. Table \ref{tab:n} depicts the power of the test in
each methodology, which shows a good performance of the proposed approach.
Nevertheless, it is not possible to compare the power between the two
methods because they have different empirical test sizes.

\centerline{[[ Table \ref{tab:n}]]}

We observe that, the results shown in Table \ref{tab:n} are similar for
other values of the parameters $\bm{a}$ and $\bm{B}$, if we maintain the same
proportionality of $\bm{\Sigma}$ and $\bm{\Sigma}_e$ as defined above. But,
our simulations suggest that the larger the measurement error, the larger
the sample size required to have a good asymptotic approximation for the
Wald-type statistics (\ref{Wald}).

We also conduct simulation studies for testing the simple hypothesis
$H_0: b_{12} = 0$ at 5\% nominal level. In this study, we keep fixed
the value of $b_{21}=0.2$. Others simulations were built considering
others values for $b_{21}$, however, the  results are close to each
other and, for this reason, we omit them. As can be seen, Tables
\ref{tab:n} and \ref{reject} present similar behaviors, i.e., the
proposed model has always empirical size test closer to the nominal
level than the usual model.

\centerline{[[ Table \ref{reject}]]}

\begin{figure}[ht!]
  \begin{center}
    \includegraphics[scale=0.75,angle=0]{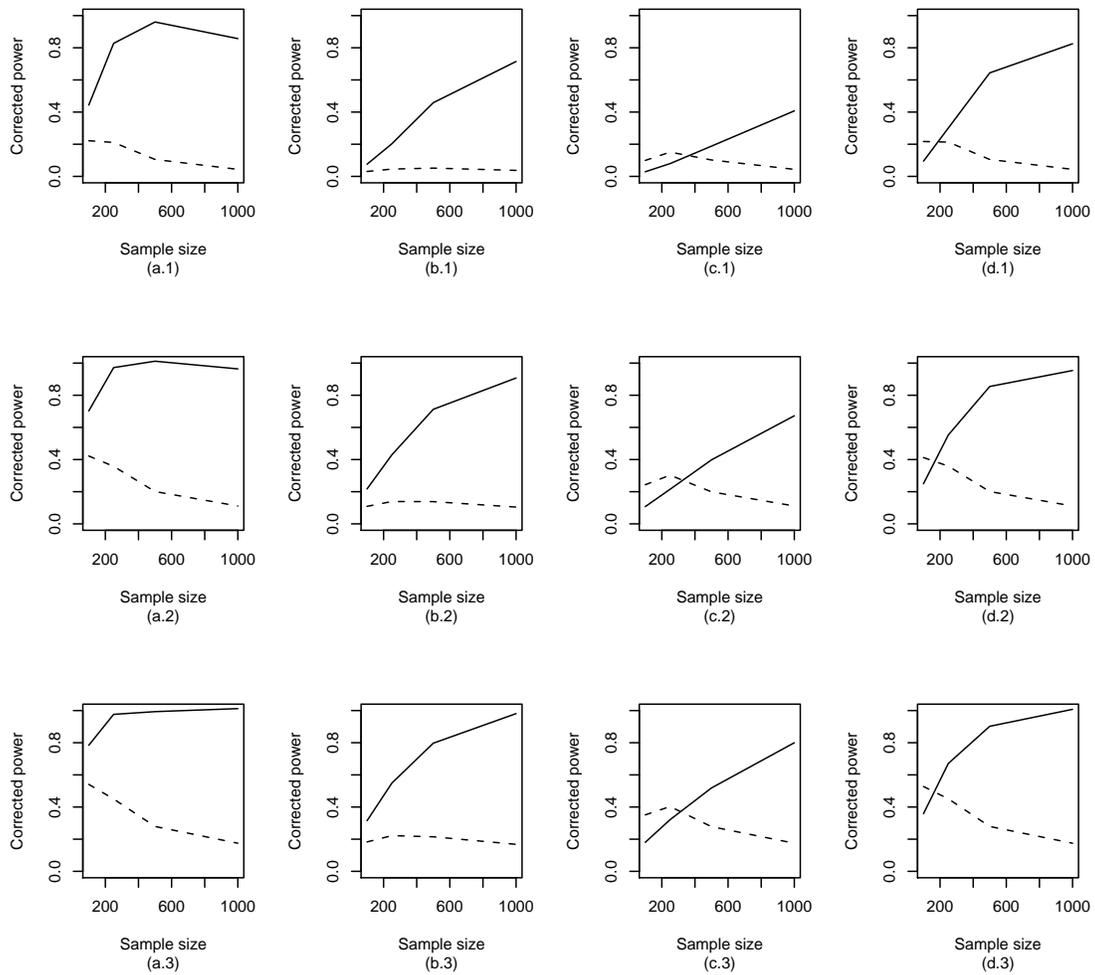}
  \end{center}
  \caption{Corrected power \emph{versus} sample size. The full line refers to the proposed approach and the
dot line refers to the usual one. It is expected that the corrected power converges to one.}
  \label{fig:1}
\end{figure}

In Table \ref{tab:n} and \ref{reject}, the usual approach seems to be
most powerful than the proposed approach when $b_{21} = 0.2$ and
$b_{21} = 0.4$. However, as aforementioned, they can not be compared
directly, just because the real nominal level used to compute that powers
 are not the same.
 Thus, we used a descriptive measure in order to analyze both
methodologies around the null hypothesis. Let
$a_n(\alpha)$ be the probability of the error type I using the true
distribution of (\ref{Wald}) when the sample size is $n$
 and $\alpha$ is the adopting nominal level based on its asymptotic distribution.
For instance, in Table \ref{reject} we have estimated $\widehat{a_{100}}(0.05)=0.0513$
for the proposed approach and $\widehat{a_{100}}(0.05) = 0.0837$ for the
usual approach. An expected behavior for good statistics is $a_n(\alpha)
\stackrel{n\rightarrow \infty}{\longrightarrow} \alpha$ which means that
the quantiles of the true distribution of (\ref{Wald}) will be close
to the quantiles of the asymptotic distribution, $\chi^2(m)$, when the sample size
is sufficiently large. Thus, the relation $a_n(\alpha)/\alpha$ tell us how far is the
$\alpha$-quantil of the asymptotic  distribution from the true distribution
of (\ref{Wald}) for each $n$. Therefore, we can define a sort of corrected power as

\[
P_n^{(c)}(\alpha) = \dfrac{ P_n(a_n(\alpha))}{(a_n(\alpha)/\alpha)}
\] where $P_n(a(\alpha))$ is the power using the true probability of the error type
I, namely $a_n(\alpha)$. We are just penalizing the power by the distance between
$a_n(\alpha)$ and $\alpha$.
Notice that, the power under the null hypothesis has to be the nominal level and for
comparing powers from different statistics it must be done using the same nominal
level. Let $a_{1n}(\alpha)$ and  $a_{2n}(\alpha)$ be the true probability of the error
type I for two different statistics when the sample size is $n$. Then, under the null hypothesis,
we have

\[ P_{1n}^{(c)}(\alpha) = P_{2n}^{(c)}(\alpha) =\alpha,
\] and hence, the corrected powers $P_{1n}^{(c)}$ and $P_{2n}^{(c)}$ are comparable.
Moreover, under an alternative hypothesis and when $n$ increases, an expected
behavior of $P_n^{(c)}(\alpha)$ is to converge to one. Although, this corrected power
is not a monotonic function of the sample size nor of the  nominal level, we believe
that it is a kind of descriptive measure to evidence how unsuitable is
the usual model when compared with the proposed one outside the null hypothesis.
 Furthermore, the proposed corrected power varies between $0$ and infinity.
 Figure \ref{fig:1} shows the corrected power for both approaches, the null hypothesis
was $H_0: b_{12} = 0$. The full line refers to the proposed
approach and the dashed line refers to the usual one. The panels  (a.1), (b.1), (c.1)
and (d.1) refer to the corrected power when the alternative hypothesis are $b_{12} = -0.4$,
$b_{12} = -0.2$, $b_{12} = 0.2$ and $b_{12} = 0.4$, respectively at $\alpha = 0.01$.
 The panels (a.2), (b.2), (c.2) and (d.2) refer to the corrected power when the
alternative hypothesis are $b_{12} = -0.4$, $b_{12} = -0.2$, $b_{12} = 0.2$ and
$b_{12} = 0.4$, respectively at $\alpha = 0.05$.
 The panels (a.3), (b.3), (c.3) and (d.3) refer to the corrected power when the
alternative hypothesis are $b_{12} = -0.4$, $b_{12} = -0.2$, $b_{12} = 0.2$ and
$b_{12} = 0.4$, respectively at $\alpha = 0.10$. We observe in all graphs that, the usual
approach has the worst performance (going to zero when the sample size increases)
while the proposed one have an expected behavior for a good statistic (going to one
when the sample size increases). In general, the corrected power under the usual methodology
goes to zero because the distance between $a_n(\alpha)$ and $\alpha$ increases much faster
 than the uncorrected power, $P_n(a_n(\alpha))$, when $n$ increases. This behavior is still
true for another setup of parameters.

\centerline{[[ Figure \ref{fig:1}]]}

\centerline{[[ Table \ref{VicioEqm}]]}

Table \ref{VicioEqm} shows that,
for this set of parameters, the biases of the estimators of $b_{ij}$ ($i,j=1,2$)
from the proposed model is smaller than the value supplied by the usual model.
Moreover, the larger the sample size, the smaller the bias and MSE under the
proposed model (this does not happen for the usual approach).

\section{\protect \centering APPLICATION}

As previously described, the models with measurement errors have great relevance in applied sciences, since equipment imprecisions are inherent to data acquisition. Actually, the usual models are commonly applied ignoring these errors. Nowadays, the scientific community started to pay enough attention to the fact that these procedures may lead to spurious results. In this section, we illustrate the concepts introduced in the present study with an application embedded in Neuroscience research, with the utilization of VAR modeling for the characterization of brain networks.

The dataset explored in this application is proceeding from a functional magnetic resonance imaging (fMRI) experiment. Basically, fMRI acquisition is based on monitoring the BOLD signal (blood oxygenation level dependent) at several brain regions through time. One of the main advantages of fMRI over other imaging techniques is its non-invasive protocol and relative high spatial resolution. The BOLD signal is related to oxygen consumption and blood flow, being considered as an indirect measure of local neural activity (\cite{logo01}). Regarding this property, this signal is used to quantify and locate the brain activity in humans.

\begin{figure}[ht!]
  \begin{center}
    \includegraphics[scale=0.75,angle=0]{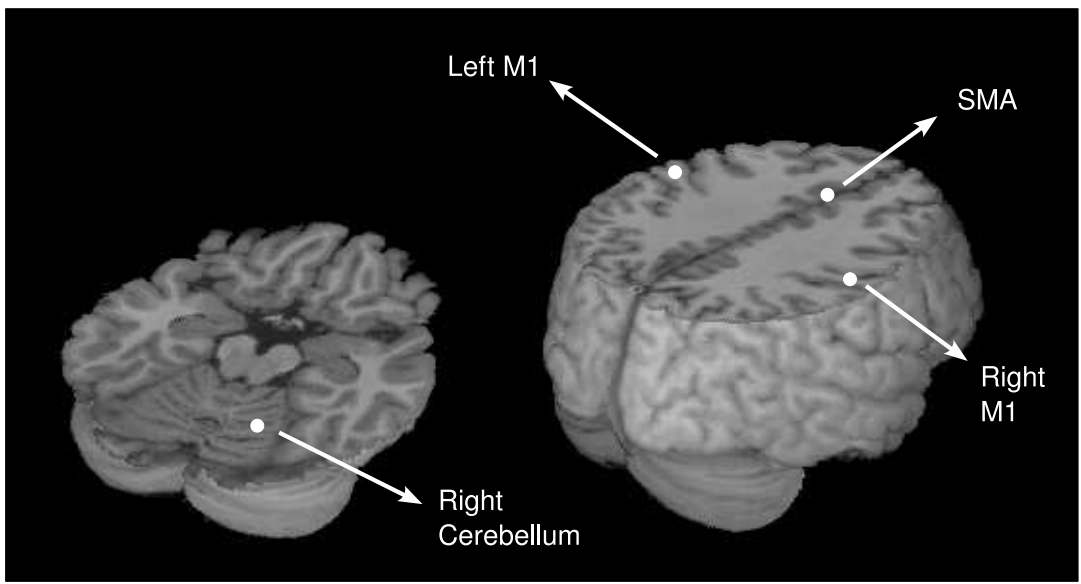}
  \end{center}
  \caption{Four areas were selected for connectivity evaluation using the VAR model: \textbf{Left M1:} left primary motor cortex, \textbf{Right M1:} right primary motor cortex, \textbf{SMA:} supplementary motor area and \textbf{Right Cerebellum.}}
  \label{figAreas}
\end{figure}

In this study, the BOLD signals at four brain regions from a subject in a resting state (eyes closed) condition were considered. The data was collected in a Siemens 3Tesla MR system (TR=1800ms, TA=900ms, TE=30ms). The selected brain regions were: left primary motor cortex (left M1), right primary motor cortex (right M1), supplementary motor area (SMA) and right cerebellum. The anatomical location of this areas is shown in Figure \ref{figAreas}. These areas are frequently involved in active and planned right hand fingertapping. We aim to evaluate the information flow between these areas in a resting state condition by using VAR models for Granger causality identification.

A well described limitation inherent to all fMRI acquisition is the high level of scanner noise. Thus, the signals observed mirror not only the physiological variations but also includes measurement errors. For this specific dataset, it was estimated that the error composed approximately $34.60\%$ of the observed time series standard deviation. For simplicity, each observed series were normalized to have mean zero and variance one. Thus, the measurement error was considered to be serially uncorrelated, independent of the latent variables and with a standard deviation of $0.346$.

The model considered for the latent variable is given by
\begin{equation}
\bm{z}_t=\bm{a}+\bm{B}_1\bm{z}_{t-1}+ \bm{q}_t,\,\,\,t=1,\cdots,n
\end{equation}
where $n=200$ is the time series length, $\bm{z}_t =(z_{1t},z_{2t},z_{3t},z_{4t})^{\top}$ with
$z_{1t}:$ the \emph{Left M1} BOLD signal, $z_{2t}:$ the \emph{SMA} BOLD signal,
$z_{3t}:$ the \emph{Right M1} BOLD signal and $z_{4t}:$ the \emph{Right cerebellum} BOLD signal;
$\mathbf{B}_1$ is the $(4\times 4)$ autoregressive coefficients matrix

\begin{eqnarray}
\mathbf{B}_1=
\begin{pmatrix}
b_{11} & b_{12} & b_{13} & b_{14}\\
b_{21} & b_{22} & b_{23} & b_{24}\\
b_{31} & b_{32} & b_{33} & b_{34}\\
b_{41} & b_{42} & b_{43} & b_{44}
\end{pmatrix},
\end{eqnarray}
and $\mathbf{q}_t$ is an $(4\times 1)$ unobservable zero mean white noise vector. The observed variables are given by

\begin{equation}
\bm{Z}_t= \bm{z}_t+\bm{e}_{t},\,\,\,t=1,\cdots,n
\end{equation}
 where $\bm{Z}_t=(Z_{1t},Z_{2t},Z_{3t},Z_{4t})^{\top}$ and $\bm{e}_{t}=(e_{1t},e_{2t},e_{3t},,e_{4t})^{\top}$ is the measurement error vector.

\begin{figure}[ht!]
  \begin{center}
    \includegraphics[scale=0.50,angle=-90]{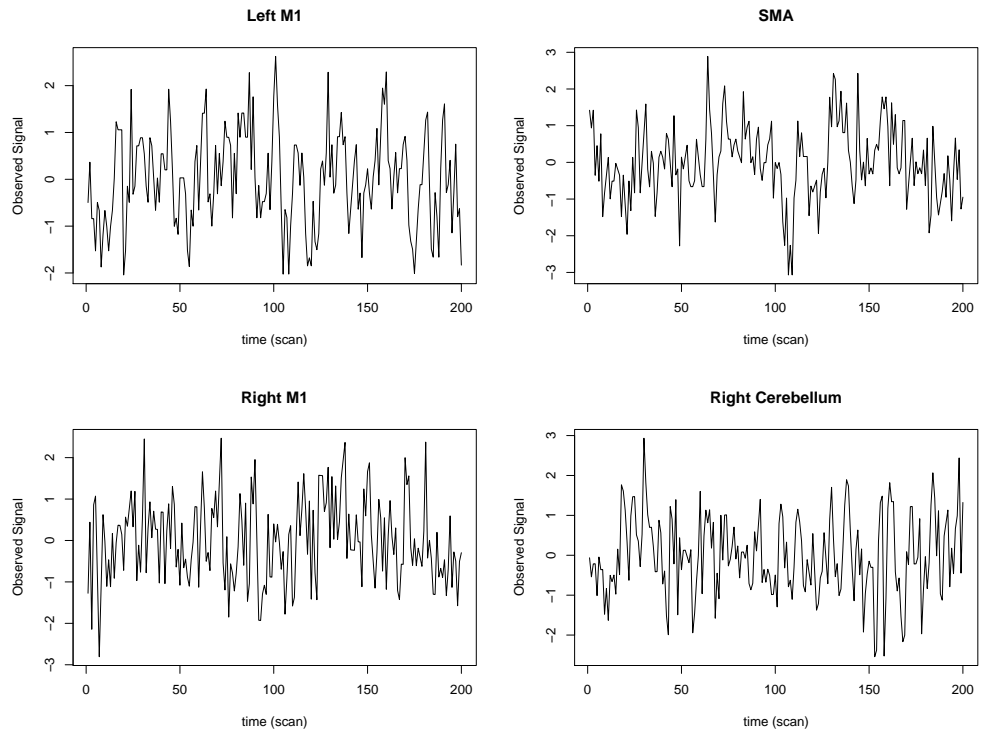}
  \end{center}
  \caption{Observed signal at each brain region.}
  \label{figSignal}
\end{figure}

The time series plots corresponding to the respective observed
BOLD signal at each brain region are represented in Figure
\ref{figSignal}. Since we are interested in identifying the links
of connectivity networks using Granger causality, the statistical
inferences are related to the parameters $b_{ij}$ $(i,j=1,2,3,4)$.
If $b_{ij}\ne0$, then there is a information flow from brain area
$j$ to area $i$ (\cite{baccala01}). The coefficient estimates,
standard errors and p-values ($H_0:b_{ij}=0$ vs $H_1:b_{ij}\ne0$)
for both usual and proposed approaches are shown in Tables
\ref{app1} and \ref{app2}, respectively.

\centerline{[[ Figure \ref{figAreas}]]}

\centerline{[[ Figure \ref{figSignal}]]}

\begin{figure}[ht!]
  \begin{center}
  \includegraphics[scale=0.60,angle=0]{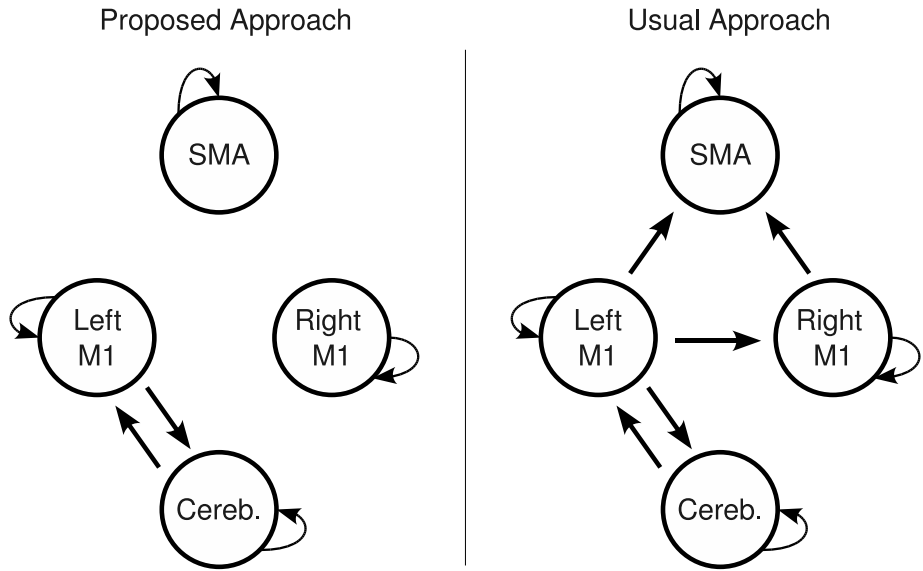}
  \end{center}
  \caption{Identified network of information flow by testing the parameters of VAR model ($\alpha=5\%$)}
  \label{figNetwork}
\end{figure}

\centerline{[[ Figure \ref{figNetwork}]]}

\begin{figure}[ht!]
  \begin{center}
    \includegraphics[scale=0.50,angle=-90]{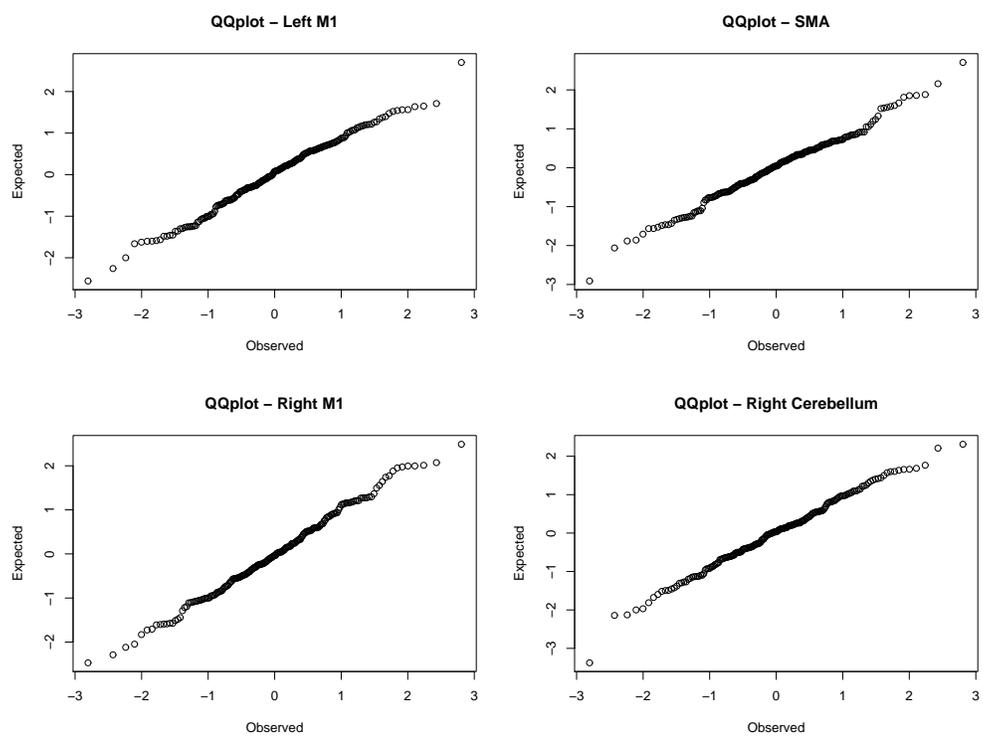}
  \end{center}
  \caption{\textbf{QQplot for Normal distribution:} Residuals (Observed values - Predicted) at each brain region.}
  \label{figQQapp}
\end{figure}

\centerline{[[ Figure \ref{figQQapp}]]}

\centerline{[[ Table \ref{app1}]]}

\centerline{[[ Table \ref{app2}]]}

The results described in Tables \ref{app1} and \ref{app2} suggest
the existence of bidirectional information flow between Left M1
and Cerebellum. However, the application of usual approach
indicates also that Left M1 sends information to SMA and Right M1,
and that the latter sends to SMA. For both usual and proposed
approaches, the diagrams of the networks at the significance level
of 5$\%$ are shown in Figure \ref{figNetwork}. As highlighted by
the simulations results, the utilization of usual VAR estimation,
ignoring the measurement errors, may result in wrong test nominal
sizes. In this context, it is important to mention that the main
differences between the usual and proposal results were on
standard deviation estimates. Further, the proposal estimates are
almost twice the values resulting from usual approach. The theory
and simulations suggest the existence of biases in the latter.
Consequently, the p-values from the usual method tend to be
underestimated, resulting in high rejection rates. Note that this
connections may possibly exist, but since the nominal level of the
test is ``incorrect'', the type I Error is not under control. In
addition, note that some coefficients were considerably
underestimated, for example $b_{11}$, $b_{22}$ and $b_{33}$.
Finally, the qq-plots represented in Figure \ref{figQQapp} suggest
that the probability density of residuals
$\bm{Z}_t-\hat{\bm{Z}}_t$ are reasonably approximated by the
Normal distribution.

Some studies (\cite{biswal95}) suggest the existence of functional networks between motor areas even in resting state condition. These studies are based on correlation analysis between the BOLD signal at different brain sites. First, it is important to note that Granger causality is conceptually different from correlation, which is symmetric (it does not provide the direction of information flow ), evaluated in a pairwise fashion (and not in the full multivariate sense) and it does not take into account temporal information. In fact, correlation analysis is more closely related to instantaneous Granger Causality concept, which can be useful to quantify simultaneity between time series but it is unsuitable in the context of information flow detection. Second, the usual correlation analysis does not consider the presence of measurement errors, which may also affect the statistical significance of results. The nature of functional networks in resting state is still unclear and is the subject of several studies (\cite{long08}). Nevertheless, we have demonstrated in this study that the inclusion of measurement errors can considerably influence the final results. Thus, the development of novel approaches dealing with this artifact is necessary.

In summary, since the proposal and usual results differ, we conclude that the presence of measurement error cannot be ignored. An adequate treatment for this artifact is essential for the adequate description and modeling of brain networks. It is surprising that this important limitation received proper attention only recently. We believe that a preliminary analysis of this problem points toward the demand for the development of new estimation procedures regarding scanner noise characterization, physiological noise and computational implementation.

\section{\protect \centering CONCLUSION}
This paper has introduced a new approach to model multivariate times series
when measurement errors are present. The simulation studies indicate
that the proposed approach gives coherent results (test size close to the
nominal level even for small samples, power increasing with the sample
size under alternative hypotheses, biases and mean square errors decreasing
when the sample size increases) under small and moderate measurement error.
Such features seem no to be shared by the conventional maximum likelihood estimators
which presents a much poorer performance. Furthermore, the proposal
is easily attained and iterative procedures are not required. The theory, simulations
and application showed that the presence of measurement error cannot be neglected and
 a proper model has to be considered for the adequate description and modeling of
brain networks. We expect to report generalizations of the proposed model (for elliptical
errors and heteroscedasticity situations), a residual study and more simulation
studies for large measurement errors  on incoming papers.

\appendix

\section{\protect \centering PROOF OF THEOREMS}

\subsection{Proof of Theorem \ref{EstHomo}}\label{A}

In order to prove the consistence of the estimators stated  in Theorem \ref{EstHomo}, namely

\[\widehat{\bm{a}}= \bar{\bm{Z}}_t-\widehat{\bm{B}}\bar{\bm{Z^*}}_{t-1}, \qquad \widehat{\bm{B}}=\bigg[(\bm{S}_{\bm{Z^*}_{t-1}}- \bm{I}_r\otimes\bm{\Sigma}_{e})^{-1}\bm{S}_{\bm{Z^*}_{t-1}\bm{Z}_t}\bigg]^{\top}
\] and

\[
\widehat{\bm{\Sigma}} =  n^{-1} \sum_{i=1}^{n} (\bm{Z}_i - \widehat{\bm{a}} - \widehat{\bm{B}}\bm{Z^*}_{i-1})(\bm{Z}_i- \widehat{\bm{a}} - \widehat{\bm{B}}\bm{Z^*}_{i-1})^{\top} -\bm{\Sigma}_e - \widehat{\bm{B}}(\bm{I}_r\otimes\bm{\Sigma}_e)\widehat{\bm{B}}^{\top},
\] we must study the limits of the quantities $\bm{S}_{\bm{Z^*}_{t-1}}$,
 $\bm{S}_{\bm{Z^*}_{t-1}\bm{Z}_t}$, $\bar{\bm{Z^*}}_{t-1}$ and $\bar{\bm{Z^*}}_{t}$
when the sample size goes to infinity. Note that $\bm{Z^*}_{t-1} = \bm{z^*}_{t-1}+ \bm{e^*}_{t-1}$,
where $\bm{e^*}_{t-1} = (\bm{e}_{t-1}^{\top}, \ldots, \bm{e}_{t-r}^{\top})^{\top}$, and
under the stationary conditions of a $\mbox{VAR}(r)$ model we have that

\begin{eqnarray*}
\bm{S}_{\bm{Z^*}_{t-1}} &=&n^{-1}\sum_{i=1}^{n}(\bm{Z^*}_{i-1}-\bar{\bm{Z^*}}_{t-1})\bm{Z}_{i-1}^{\bm{*}\top}\\
&=& n^{-1}\sum_{i=1}^{n}(\bm{z^*}_{i-1} + \bm{e^*}_{i-1}-\bar{\bm{z^*}}_{t-1} - \bar{\bm{e^*}}_{t-1})(\bm{z^*}_{i-1} + \bm{e^*}_{i-1})^{\top}\\
&=& \bm{S}_{\bm{z^*}_{t-1}}  + \bm{S}_{\bm{e^*}_{t-1}}  + O_p(n^{-1/2})\\
&=& \bm{\Gamma}_r(0)  + \bm{I}_r \otimes \bm{\Sigma}_e + O_p(n^{-1/2}),
\end{eqnarray*}

\noindent where $\bm{S}_{\bm{e^*}_{t-1}} = n^{-1}\sum_{i=1}^{n}\bm{e^*}_{i-1}\bm{e}_{i-1}^{\bm{*}\top}$, and
$O_p(n^{-1/2})$ means limited in probability even multiplying by $n^{1/2}$ (it happens with the crossing product
in the above expression). That is,
$\bm{S}_{\bm{Z^*}_{t-1}} \stackrel{\mathcal{P}}{\longrightarrow} \bm{\Gamma}_r(0)  + \bm{I}_r \otimes \bm{\Sigma}_e$.
Following the same scheme, we have that

\begin{eqnarray*}
\bm{S}_{\bm{Z^*}_{t-1}\bm{Z}_t} &=&n^{-1}\sum_{i=1}^{n}(\bm{Z^*}_{i-1}-\bar{\bm{Z^*}}_{t-1})\bm{Z}_{i}^{\top}\\
&=& n^{-1}\sum_{i=1}^{n}(\bm{z^*}_{i-1} + \bm{e^*}_{i-1}-\bar{\bm{z^*}}_{t-1} - \bar{\bm{e^*}}_{t-1})(\bm{z}_{i} + \bm{e}_{i})^{\top}\\
&=& \bm{S}_{\bm{z^*}_{t-1}\bm{z}_t}   + O_p(n^{-1/2})\\
&=& \bm{\Gamma}_r(0)\bm{B}^{\top} + O_p(n^{-1/2}),
\end{eqnarray*} and finally, both the quantities $\bar{\bm{Z^*}}_{t-1}$ and $\bar{\bm{Z^*}}_{t}$
converge in probability to $ \bm{\mu^*}$. Hence,

\[
(\bm{S}_{\bm{Z^*}_{t-1}}-  \bm{I}_r \otimes \bm{\Sigma}_e) ^{-1} \stackrel{\mathcal{P}}{\longrightarrow} \bm{\Gamma}_r(0)^{-1}  \qquad \mbox{and} \qquad \bm{S}_{\bm{Z^*}_{t-1}\bm{Z}_t}  \stackrel{\mathcal{P}}{\longrightarrow}\bm{\Gamma}_r(0)\bm{B}^{\top},
\] thus, the probability convergence of $\widehat{\bm{a}}$,
$\widehat{\bm{B}}$ and $\widehat{\bm{\Sigma}}$ to $\bm{a}$, $\bm{B}$ and $\bm{\Sigma}$ follow,
respectively.

\subsection{Proof of Theorem \ref{AsympEstHomo}}\label{B}
The proof idea has three steps. The first step consists in show that
$\mbox{vec}(\widehat{\bm{B}}^{\top}) - \mbox{vec}(\bm{B}^{\top})$ can be written
as linear combinations of a vectorial mean. The second one, we must demonstrate
that this vectorial mean has an asymptotic normal distribution. The last step must
conclude that $\mbox{vec}(\widehat{\bm{B}}^{\top}) - \mbox{vec}(\bm{B}^{\top})$
also has an asymptotic normal distribution. In order to prove Theorem \ref{AsympEstHomo},
we need some auxiliary results, which are exposed in two propositions below.

\begin{proposition}\label{Prop1} Under the model (\ref{var}) and (\ref{erro}), the proposed estimator $\widehat{\bm{B}}$ has the following relationship

\begin{equation}\nonumber
\mbox{vec}(\widehat{\bm{B}}^{\top}) - \mbox{vec}(\bm{B}^{\top}) = (\bm{I}_p\otimes\bm{\Gamma}_r(0)^{-1})\bar{\bm{W}}
+O_p(n^{-1}),
\end{equation} where
\[\bar{\bm{W}}=n^{-1}\sum_{i=1}^n\left(\begin{array}{c}
    \bm{W}_{1i}\\
    \vdots\\
    \bm{W}_{qi}\\
\end{array}\right) = n^{-1}\sum_{i=1}^n \bm{W}_i\] with
$\bm{W}_i = (\bm{q}_i+\bm{e}_{i}-\bm{B}\bm{e^*}_{i-1})\otimes(\bm{z^*}_{i-1}-\bm{\mu^*} + \bm{e^*}_{i-1})-\bm{\Psi}$ and
$\bm{\Psi} = [\bm{I}_p \otimes (\bm{I}_r\otimes \bm{\Sigma}_e)]\mbox{vec}(\bm{B}^{\top})$.
\end{proposition}

\noindent {\bf Proof:} Define $\bm{B}_{.k}$ as  a vector ($rp\times 1$) of coefficients associated with the $k^{th}$
element of the vector $\bm{z}_{t}$, that is
\[
z_{kt} = a_k + \bm{B}_{.k}^{\top} \bm{z^*}_{t-1} + q_{kt}.
\] Thus, we have that
$\mbox{vec}(\bm{B}^{\top})=({\bm{B}}_{.1}^{\top},{\bm{B}}_{.2}^{\top},\cdots,{\bm{B}}_{.p}^{\top})^{\top}$ and
the estimator of Theorem \ref{EstHomo} for it can be written as
$\mbox{vec}(\widehat{\bm{B}})=(\widehat{\bm{B}}_{.1}^{\top},\widehat{\bm{B}}_{.2}^{\top},\cdots,\widehat{\bm{B}}_{.p}^{\top})^{\top}$,
where $\widehat{\bm{B}}_{.k}=(\bm{S}_{\bm{Z^*}_{t-1}}-\bm{I}\otimes\bm{\Sigma}_{e})^{-1}\bm{S}_{\bm{Z^*}_{t-1}Z_{kt}}$
and $\bm{S}_{\bm{Z^*}_{t-1}Z_{kt}}=n^{-1}\sum_{i=1}^{n}(\bm{Z^*}_{i-1}-\bar{\bm{Z^*}}_{t-1})Z_{kt}$
for $k = 1,\ldots, p$. Moreover, the model (\ref{varCOL}) may be rewritten in terms of the observed variables as

\begin{equation}\label{ev}
\begin{array}{ccc}
\bm{Z}_t & = & \bm{a}+\bm{B} \bm{Z^*}_{t-1} + \bm{\vartheta}_t,\\
\bm{\vartheta}_t & = & \bm{q}_t + \bm{e}_{t} - \bm{B}\bm{e^*}_{t-1},
\end{array}
\end{equation} and for the $k^{th}$ element of $\bm{Z}_t$ we have

\begin{equation}\label{evK}
\begin{array}{ccc}
Z_{kt} & = & a_k+ \bm{B}_{.k}^{\top} \bm{Z^*}_{t-1} + \vartheta_{kt},\\
\vartheta_{kt} & = & q_{kt} + e_{kt} - \bm{B}_{.k}^{\top}\bm{e^*}_{t-1}.
\end{array}
\end{equation}

\noindent Then, it follows that

\[\bm{S}_{\bm{Z^*}_{t-1}Z_k}=n^{-1}\sum_{i=1}^n(\bm{Z^*}_{i-1}- \bar{\bm{Z^*}}_{t-1})(a_k+\bm{B}_{.k}^{\top}\bm{Z^*}_{i-1}+\vartheta_{ki})=\bm{S}_{\bm{Z^*}_{t-1}}\bm{B}_{.k} + \bm{S}_{\bm{Z^*}_{t-1}\vartheta_k},
\] where $\bm{S}_{\bm{Z^*}_{t-1}\vartheta_k}=n^{-1}\sum_{i=1}^n(\bm{Z^*}_{i-1}- \bar{\bm{Z^*}}_{t-1})\vartheta_{ki} =
n^{-1}\sum_{i=1}^n(\bm{z^*}_{i-1}- \bm{\mu^*} + \bm{e^*}_{i-1})\vartheta_{ki} + O_p(n^{-1})$. Hence,
denoting $\bm{S}_{\bm{z^*}_{t-1}\vartheta_k} = n^{-1}\sum_{i=1}^n(\bm{z^*}_{i-1}- \bm{\mu^*} + \bm{e^*}_{i-1})\vartheta_{ki}$ we have that

\[\bm{S}_{\bm{Z^*}_{t-1}Z_k}= (\bm{S}_{\bm{Z^*}_{t-1}} - \bm{I}_r \otimes \bm{\Sigma}_e)\bm{B}_{.k} + \bm{S}_{\bm{z^*}_{t-1}\vartheta_k} - \bm{\Psi}_k+ O_p(n^{-1}),
\] with $\bm{\Psi}_k = -\bm{I}_r \otimes \bm{\Sigma}_e\bm{B}_{.k}$. As a result, we have

\[
\widehat{\bm{B}}_{.k}=\bm{B}_{.k}+\bm{\Gamma}_r^{-1}(0)\bar{\bm{W}}_k+O_p(n^{-1})
\] where $\bar{\bm{W}}_k = n^{-1}\sum_{i=1}^n \bm{W}_{ki}$ and
$\bm{W}_{ki} = (\bm{z^*}_{i-1}- \bm{\mu^*} + \bm{e^*}_{i-1})\vartheta_{ki} - \bm{\Psi}_k$. Hence, it
follows that

\begin{equation}\nonumber
\mbox{vec}(\widehat{\bm{B}}^{\top}) - \mbox{vec}(\bm{B}^{\top}) = (\bm{I}_p\otimes\bm{\Gamma}_r(0)^{-1})\bar{\bm{W}}
+O_p(n^{-1}),
\end{equation} where
\[\bar{\bm{W}}=n^{-1}\sum_{i=1}^n\left(\begin{array}{c}
    \bm{W}_{1i}\\
    \vdots\\
    \bm{W}_{qi}\\
\end{array}\right) = n^{-1}\sum_{i=1}^n \bm{W}_i\] with
$\bm{W}_i = (\bm{q}_i+\bm{e}_{i}-\bm{B}\bm{e^*}_{i-1})\otimes(\bm{z^*}_{i-1}-\bm{\mu^*} + \bm{e^*}_{i-1})-\bm{\Psi}$ and
$\bm{\Psi} = [\bm{I}_p \otimes (\bm{I}_r\otimes \bm{\Sigma}_e)]\mbox{vec}(\bm{B}^{\top})$.

\begin{proposition} \label{Prop2} If $\bm{e}_{t} \sim \mathcal{N}(\bm{0}, \bm{\Sigma}_e)$ with
$\bm{\Sigma}_{e}$ known and $\mbox{E}(q_{ij_1}q_{ij_2}q_{ij_3}q_{ij_4}) < \infty$
for all $j_1, j_2, j_3, j_4 \in \{1,\ldots, p\}$, where $q_{ij}$ is the $j^{th}$ element of
$\bm{q}_i$. The mean, $\bar{\bm{W}}$, of Proposition \ref{Prop1} has an asymptotic distribution
given by

\[
\sqrt{n} \bar{\bm{W}} \stackrel{\mathcal{D}}{\longrightarrow} \mathcal{N}(\bm{0}, \bm{T}_r),
\] where

\begin{eqnarray*}
\bm{T}_r& =& \bm{\Sigma}_{\vartheta}\otimes \bm{\Gamma}_r(0) + \bm{\Sigma}_{\vartheta} \otimes (\bm{I}_r \otimes \bm{\Sigma}_e) + \bm{B}^{\top} \otimes  [\bm{\Sigma}_e\bm{B} (\bm{I}_r \otimes \bm{\Sigma}_e)] +\\
&&  - \sum_{h=1}^{r}\bigg\{( \bm{B}_{h} \bm{\Sigma}_{e})\otimes\bm{\Gamma}_r(h) + (\bm{\Sigma}_{e}\bm{B}_{h}^{\top})\otimes\bm{\Gamma}_r(-h)\bigg\} +\\
&& +  \sum_{h=1-r}^{r-1} [\bm{B}(\bm{J}_{-h} \otimes \bm{\Sigma}_e)\bm{B}^{\top}]\otimes \bm{\Gamma}_r(h).
\end{eqnarray*}
where $\bm{J}_{l}$ is a ($r \times r$) matrix of zeros with one's in the $|l|^{th}$ diagonal above (below) the main diagonal if $l>0$ ($l<0$) and $\bm{J}_0$ is a ($r \times r$) matrix of zeros.
\end{proposition}

\noindent {\bf Proof:} Notice that the expectation of $\bm{W}_i$ is equal to zero for all $i$.
\cite{Shumway} state a central limit theorem
to a univariate M-dependent sequence of random variables with mean zero. We say that a time series $x_t$  is
M-dependent if the set of values $x_s, s\leq t$ is independent of the set of values $x_s, s\geq t+M+1$
\citep[][on pg. 66]{Shumway}. Then, assuming that $\mbox{E}(q_{ij_1}q_{ij_2}q_{ij_3}q_{ij_4})< \infty$
for all $j_1, j_2, j_3, j_4 \in \{1,\ldots,p\}$ where $q_{ij}$ is the $j^{th}$ element of $\bm{q}_i$
and defining $\bar{x} = n^{-1}\sum_{i=1}^n x_i$, where $x_i = \bm{\delta}^{\top}\bm{W}_i$ we have that
$\mbox{E}(x_i)=0$, $\mbox{Cov}(x_i,x_{i-h}) = \bm{\delta}^{\top}\mbox{Cov}(\bm{W}_i,\bm{W}_{i-h}^{\top})\bm{\delta} = \bm{\delta}^{\top}E(\bm{W}_i\bm{W}_{i-h}^{\top})\bm{\delta} $ and

\begin{eqnarray*}
E(\bm{W}_i\bm{W}_{i-h}^{\top})  &=& E[\bm{F}_{ih} \otimes(\bm{z^*}_{i-1}-\bm{\mu^*})(\bm{z^*}_{i-h-1}-\bm{\mu^*})^{\top}] +E[\bm{F}_{ih} \otimes \bm{e^*}_{i-1}\bm{e}_{i-h-1}^{\bm{*}\top}] +\\
&+& E[\bm{F}_{ih} \otimes \bm{e^*}_{i-1} (\bm{z^*}_{i-h-1}-\bm{\mu^*})^{\top}] +E[\bm{F}_{ih} \otimes(\bm{z^*}_{i-1}-\bm{\mu^*})\bm{e}_{i-h-1}^{\bm{*}\top}] - \\
&-&\bm{\Psi}\bm{\Psi}^{\top}
\end{eqnarray*} with $\bm{F}_{ih} = (\bm{q}_i+\bm{e}_{i}-\bm{B}\bm{e^*}_{i-1})(\bm{q}_{i-h}+\bm{e}_{i-h}-\bm{B}\bm{e^*}_{i-h-1})^{\top}$.
Thus, using some matricial results and simple expectation rules we can solve these expectations as follows

\[E(\bm{W}_i\bm{W}_{i-h}^{\top}) = \bm{0} \qquad \mbox{for} \quad |h|<r,\]
\[E(\bm{W}_i\bm{W}_{i-h}^{\top})=-(\bm{B}_{r} \bm{\Sigma}_{e})\otimes\bm{\Gamma}_r(h) \qquad \mbox{for}\quad  h=r,\]
\[E(\bm{W}_i\bm{W}_{i-h}^{\top})=-(\bm{\Sigma}_{e}\bm{B}_{|r|}^{\top})\otimes\bm{\Gamma}_r(h) \qquad \mbox{for}\quad  h= -r,\]
\[E(\bm{W}_i\bm{W}_{i-h}^{\top})=[\bm{B}(\bm{J}_{-h} \otimes \bm{\Sigma}_e)\bm{B}^{\top}]\otimes \bm{\Gamma}_r(h) - (\bm{B}_{h} \bm{\Sigma}_{e})\otimes\bm{\Gamma}_r(h) \qquad \mbox{for} \quad h = 1,\ldots, r-1,\]
\[E(\bm{W}_i\bm{W}_{i-h}^{\top})=[\bm{B}(\bm{J}_{-h} \otimes \bm{\Sigma}_e)\bm{B}^{\top}]\otimes \bm{\Gamma}_r(h) - ( \bm{\Sigma}_{e}\bm{B}_{|h|}^{\top})\otimes\bm{\Gamma}_r(h) \qquad \mbox{for} \quad h = -1,\ldots, 1-r,\]
\[
E(\bm{W}_i\bm{W}_{i-h}^{\top}) = \bm{\Sigma}_{\vartheta}\otimes \bm{\Gamma}_r(0) + \bm{\Sigma}_{\vartheta} \otimes (\bm{I}_r \otimes \bm{\Sigma}_e) + \bm{B}^{\top} \otimes  [\bm{\Sigma}_e\bm{B} (\bm{I}_r \otimes \bm{\Sigma}_e)] \qquad \mbox{for} \quad h=0,
\] where $\bm{J}_{l}$ is a ($r \times r$) matrix of zeros with one's in the $|l|^{th}$ diagonal above (below) the main diagonal if $l>0$ ($l<0$) and $\bm{J}_0$ is a ($r \times r$) matrix of zeros.
That is, $x_1\ldots, x_n$ is a strictly M-dependent sequence of random variables with mean zero (where $M=r$) and,
therefore, we can use the result stated in \cite{Shumway}, which says that
\[
\sqrt{n}\bar{x} \stackrel{\mathcal{D}}{\longrightarrow} \mathcal{N}(0, V_r)
\] where

\[
V_r = \sum_{h=-r}^{r} \mbox{Cov}(\bm{\delta}^{\top}\bm{W}_i,\bm{\delta}^{\top}\bm{W}_{i-h}) = \bm{\delta}^{\top}\bm{T}_r\bm{\delta}\] with

\begin{eqnarray*}
\bm{T}_r& =& \bm{\Sigma}_{\vartheta}\otimes \bm{\Gamma}_r(0) + \bm{\Sigma}_{\vartheta} \otimes (\bm{I}_r \otimes \bm{\Sigma}_e) + \bm{B}^{\top} \otimes  [\bm{\Sigma}_e\bm{B} (\bm{I}_r \otimes \bm{\Sigma}_e)] +\\
&&  - \sum_{h=1}^{r}\bigg\{( \bm{B}_{h} \bm{\Sigma}_{e})\otimes\bm{\Gamma}_r(h) + (\bm{\Sigma}_{e}\bm{B}_{h}^{\top})\otimes\bm{\Gamma}_r(-h)\bigg\} +\\
&& +  \sum_{h=1-r}^{r-1} [\bm{B}(\bm{J}_{-h} \otimes \bm{\Sigma}_e)\bm{B}^{\top}]\otimes \bm{\Gamma}_r(h).
\end{eqnarray*}

As $\sqrt{n}\bm{\delta}^{\top} \bar{\bm{W}}$ is asymptotically normally distributed for all $\bm{\delta} \neq \bm{0}_r$ then,
by the Cramer-Wold device \citep[see Theorem 10.4.5 on page 336 in][]{Athreya}, we have that

\[
\sqrt{n} \bar{\bm{W}} \stackrel{\mathcal{D}}{\longrightarrow} \mathcal{N}(\bm{0}, \bm{T}_r).
\]

Then, by the Propositions \ref{Prop1} and \ref{Prop2}, the prove of Theorem \ref{AsympEstHomo} follows

\[
\sqrt{n} (\mbox{vec}(\widehat{\bm{B}}^{\top}) - \mbox{vec}({\bm{B}}^{\top})) \stackrel{\mathcal{D}}{\longrightarrow} \mathcal{N}(\bm{0},  [\bm{I}_p \otimes \bm{\Gamma}_r(0)^{-1}]\bm{T}_r [\bm{I}_p \otimes \bm{\Gamma}_r(0)^{-1}]).
\]

\newpage

\renewcommand{\baselinestretch}{1.3}

\begin{table}[ht!]\caption{Rejection rates (\%) of the hypothesis $H_0 : b_{12} =b_{21} = 0$
(at 5\% nominal level) using the Wald statistics (\ref{Wald}) for
$n=50$, $n=100$, $n=250$ and $n=500$. The bold numbers at the center
are test sizes (they are expected to be 5\%) and the numbers around them are empirical powers.} \label{tab:n}
  \begin{center}
    \begin{tabular}{lrrrrrr|rrrrr}\hline
    &     &\multicolumn{5}{c}{Corrected approach} & \multicolumn{5}{c}{Usual approach (OLS)}\\\cline{3-12}
    &     &       &      &$b_{12}$&    &      &       &     & $b_{12}$&   &  \\
    &     &  -0.4 & -0.2 &  0.0 & 0.2  &  0.4 & -0.4 & -0.2 & 0.0  &  0.2 &  0.4 \\\cline{1-12}
    &$n=50$& &&&&&&&&&\\
    &-0.4 & 29.1 & 12.6 &7.8& 11.0& 21.1& 17.5& 7.7& 10.3& 21.1& 38.7\\
    &-0.2 & 25.0 & 8.8 &5.3& 8.2& 17.2& 15.1& 6.8& 10.5& 22.6& 40.8\\
$b_{21}$&0.0  & 23.6 & 8.3 &{\bf 4.5}&7.7& 16.1& 16.1& 8.4& {\bf 13.3}& 28.1& 47.0\\
    &0.2  & 24.4 & 8.5 &5.2& 9.0& 17.3& 18.8& 11.0& 17.8& 35.4& 54.9\\
    &0.4  & 28.6 & 10.5 &7.3& 10.9& 21.6& 25.2& 15.6& 24.0& 43.5& 65.9\\\hline
    &$n=100$& &&&&&&&&&\\
    &-0.4& 59.1& 24.8 &12.2& 16.5& 34.9& 33.2& 9.4& 12.4&   34.5&63.8\\
    &-0.2& 50.5& 16.0 &6.5& 10.3& 27.4& 27.2& 6.6& 13.3&   37.4& 66.8\\
$b_{21}$&0.0 & 45.2& 12.9 &{\bf 5.2}& 10.5&27.2& 26.4& 8.5& {\bf 19.2}&48.6& 76.5\\
    &0.2 & 44.5& 12.7 &6.4& 14.0& 33.4& 31.4& 13.1& 28.7&   61.7& 86.3\\
    &0.4 & 50.3& 16.6 &10.2& 20.3& 42.2& 41.7& 22.7& 41.4&   74.3& 92.9\\\hline
    &$n=250$& &&&&&&&&&\\
    &-0.4& 95.2& 58.5 &25.8& 32.2& 68.9& 71.8& 15.9& 21.1& 67.5& 95.7\\
    &-0.2& 89.9& 37.0 &9.0& 18.5& 57.7& 59.9& 6.9& 23.9& 74.2& 97.4\\
$b_{21}$&0.0 & 84.7& 25.1 &{\bf 5.3}& 19.2& 60.6&57.7& 10.3& {\bf 39.5}& 86.8& 99.1\\
    &0.2 & 84.9& 23.5 &8.6& 29.6& 71.3& 67.4& 23.6& 60.9& 95.8& 99.9\\
    &0.4 & 87.7& 31.6 &19.6& 47.5&83.4 & 80.2& 45.9& 80.6& 98.8& 100.0\\\hline
    &$n=500$& &&&&&&&&&\\
    &-0.4& 99.9 &88.2& 47.6&  56.6&93.5& 95.7& 27.1& 37.9& 93.4 & 100.0\\
    &-0.2& 99.7 &66.0& 13.4&  32.4&87.8& 89.4& 8.3& 40.6& 96.4 & 100.0\\
$b_{21}$&0.0 & 99.2 &46.3& {\bf 5.2}&  34.0&89.9&87.8& 14.8& {\bf 66.8}& 99.3 & 100.0\\
    &0.2 & 99.0 &42.3& 12.1 & 53.9&95.9& 93.5& 42.3& 88.9& 99.9 & 100.0\\
    &0.4 & 99.4 &55.7& 36.0 & 79.0&99.0& 98.2& 76.2& 98.2& 100.0 & 100.0\\\hline
    \end{tabular}
  \end{center}
\end{table}


\begin{table}[ht]\caption{Rejection rates (\%) of the hypothesis
$H_0 : b_{12} = 0$ (at 5\% nominal level) using the Wald statistics
(\ref{Wald}) for $n=50$, $n=100$, $n=250$ and $n=500$.  } \label{reject}
  \begin{center}
    \begin{tabular}{lccccc}\hline
         &\multicolumn{5}{c}{$b_{12}$}\\
 Model    &$-0.4$   &$-0.2$  &$0.0$ &  $0.2$ & $0.4$ \\\cline{2-6}\hline
$n=50$&&&&\\
Proposed Model &    43.80  &  14.43  & {\bf 5.27} &   6.95 &  13.55\\
Usual Model    &    36.79  &  10.16  & {\bf 6.94} &  17.49 &  33.24\\\hline
$n=100$&&&&\\
Proposed Model &  71.65&   21.91& {\bf 5.13}&  10.79&  25.17\\
Usual Model   &  61.90&   12.39&  {\bf 8.37}&  32.27&  60.75\\\hline
$n=250$&&&&\\
Proposed Model&  97.58 &  43.04&  {\bf 5.02}&  21.53&  55.45\\
Usual  Model  &  94.43&   21.52&  {\bf 13.71}& 68.83&  95.32\\\hline
$n=500$&&&&\\
Proposed Model &  99.96&   70.57& {\bf 4.94}&  39.51&  84.58\\
Usual Model&      99.88&   37.49& {\bf 24.89}& 94.32&  99.95\\\hline
    \end{tabular}
  \end{center}
\end{table}

\begin{table}[ht!]\caption{Empirical bias and mean squared error for the proposed and usual model.
Note that, the biases} \label{VicioEqm}
  \begin{center}
    \begin{tabular}{lcccc}\hline
  &\multicolumn{2}{c}{Proposed model} & \multicolumn{2}{c}{Usual model}\\
  & Bias   & MSE & Bias& MSE\\\hline
$n=50$&&&&\\
$b_{11}$ &-0.0100& 0.0459 &-0.1446 & 0.0454 \\
$b_{12}$ &-0.0647& 0.0809 & 0.1098 & 0.0434 \\
$b_{21}$ & 0.0290& 0.0269 & 0.0250 & 0.0143 \\
$b_{22}$ & -0.0765& 0.0472& -0.1589&  0.0461\\\hline
$n=100$&&&&\\
$b_{11}$& -0.0035& 0.0203 &-0.1313 &0.0290\\
$b_{12}$& -0.0335& 0.0328 & 0.1209 &0.0293\\
$b_{21}$&  0.0127& 0.0115 & 0.0165 &0.0067\\
$b_{22}$& -0.0343& 0.0183 &-0.1265 &0.0258\\\hline
$n=250$&&&&\\
$b_{11}$& -0.0022& 0.0075 &-0.1252 &0.0203\\
$b_{12}$& -0.0118& 0.0112 & 0.1299 &0.0224\\
$b_{21}$& 0.0040& 0.0043  &0.0112 &0.0027\\
$b_{22}$& -0.0128& 0.0063 &-0.1086 &0.0156\\\hline
$n=500$&&&&\\
$b_{11}$& -0.0019& 0.0037 &-0.1235 &0.0175\\
$b_{12}$& -0.0053& 0.0054 & 0.1326 &0.0203\\
$b_{21}$&  0.0018& 0.0021 & 0.0097 &0.0013\\
$b_{22}$& -0.0057& 0.0030 &-0.1024 &0.0124\\\hline
    \end{tabular}
  \end{center}
\end{table}


\begin{table}[ht!]\caption{\textbf{Application to real data - usual approach:} coefficient estimates, standard deviations and respective p-values ($H_0 :$ coefficient is equal to zero).}\label{app1}
\begin{center}
    \begin{tabular}{cccc}\hline
\textbf{Parameter} & \textbf{Estimate} & \textbf{Standard Deviation} & \textbf{p$-$value}\\
\hline\hline
$b_{11}$ & 0.537 &0.065 &$<$0.001\\
$b_{12}$ & 0.105 &0.063 &0.097\\
$b_{13}$ & 0.003 &0.060 &0.967\\
$b_{14}$ &-0.181 &0.059 &0.002\\
$b_{21}$ & 0.179 &0.068 &0.008\\
$b_{22}$ & 0.378 &0.066 &$<$0.001\\
$b_{23}$ & 0.145 &0.063 &0.002\\
$b_{24}$ & 0.047 &0.062 &0.442\\
$b_{31}$ & 0.165 &0.076 &0.030\\
$b_{32}$ &-0.074 &0.074 &0.319\\
$b_{33}$ & 0.242 &0.071 &$<$0.001\\
$b_{34}$ &-0.061 &0.069 &0.378\\
$b_{41}$ & 0.294 &0.070 &$<$0.001\\
$b_{42}$ &-0.060 &0.068 &0.381\\
$b_{43}$ & 0.092 &0.065 &0.154\\
$b_{44}$ & 0.350 &0.064 &$<$0.001\\
\hline
  \end{tabular}
  \end{center}
\end{table}

\begin{table}[ht!]\caption{\textbf{Application to real data - proposed approach:} coefficient estimates, standard deviations and respective p-values ($H_0 :$ coefficient is equal to zero).}\label{app2}
  \begin{center}
    \begin{tabular}{cccc}\hline
\textbf{Parameter} & \textbf{Estimate} & \textbf{Standard Deviation} & \textbf{p$-$value}\\
\hline\hline
$b_{11}$ & 0.935 &0.137 &$<$0.001\\
$b_{12}$ &-0.032 &0.127 &0.803\\
$b_{13}$ &-0.095 &0.103 &0.357\\
$b_{14}$ &-0.287 &0.091 &0.002\\
$b_{21}$ & 0.132 &0.137 &0.332\\
$b_{22}$ & 0.581 &0.126 &$<$0.001\\
$b_{23}$ & 0.199 &0.103 &0.053\\
$b_{24}$ & 0.027 &0.092 &0.765\\
$b_{31}$ & 0.279 &0.156 &0.073\\
$b_{32}$ &-0.184 &0.143 &0.201\\
$b_{33}$ & 0.346 &0.117 &0.004\\
$b_{34}$ &-0.111 &0.106 &0.294\\
$b_{41}$ & 0.538 &0.147 &$<$0.001\\
$b_{42}$ &-0.252 &0.135 &0.063\\
$b_{43}$ & 0.044 &0.110 &0.687\\
$b_{44}$ & 0.528 &0.099 &$<$0.001\\
\hline
  \end{tabular}
  \end{center}
\end{table}

\end{document}